HOW TO PRINT THIS PAPER:
1. Split this file into a TeX code file + the uuencoded figures file. 
The separation between the files is indicated by the keyword 
"FIGURES" in capitals.
2. Latex the code file (RevTeX 3.0). The macros file "epsf.tex" is
needed for the figures to print from the TeX document.
DELETE UP TO NEXT LINE
\documentstyle[preprint,prb,aps]{revtex}
\begin{document}
\draft
\title{Semi-classical description of the frustrated antiferromagnetic chain}
\author{D. Allen and D. S\'en\'echal}
\address{Centre de Recherche en Physique du Solide et 
D\'epartement de Physique,}
\address{Universit\'e de Sherbrooke, Sherbrooke, Qu\'ebec, Canada J1K 2R1.}
\date{September 1994}
\maketitle
\begin{abstract}
The antiferromagnetic Heisenberg model on a chain with nearest and
next nearest neighbor couplings is mapped onto the $SO(3)$ nonlinear sigma
model in the continuum limit. In one spatial dimension this model is always
in its disordered phase and a gap opens to excited states. The latter form a
doubly degenerate spin-1 branch at all orders in $1/N$. We argue that this
feature should be present in the spin-1 Heisenberg model itself. Exact
diagonalizations are used to support this claim. The inapplicability of this
model to half-integer spin chains is discussed. 
\end{abstract}
\section{Introduction}
Interest in low-dimensional magnetic systems has been great in recent years,
partly because of the widespread belief that magnetism plays a key role in
high-temperature superconductivity, but also because of the successful
application of field-theoretic methods to these systems, in particular to
spin chains. 
Indeed, a mapping from the spin-$s$ antiferromagnetic (AF)
Heisenberg chain with nearest-neighbor (NN) coupling to the $O(3)/O(2)$ sigma
model has led Haldane to conjecture\cite{Haldane83} that integer spin chains
should exhibit a gap to a triplet of excited states, whereas half-integer
spin chains should not. This conjecture has later been confirmed by numerical
calculations.\cite{Nightingale86,White92}
Such a mapping may be termed {\it semi-classical}, since it is
constructed by introducing a local field characterizing the order expected
in the classical ground state and by taking into account fluctuations of the
spin variables around this local order. In the case of the simple AF chain,
a collinear order exists in the classical ground state and the local field
introduced is a unit vector ${\bf e}(x,t)$ representing the staggered
magnetization. The long wavelength effective action obtained for this
field is that of the $O(3)/O(2)$ sigma model, with Lagrangian density
\begin{equation}
\label{NLs}
{\cal L}_\sigma = {1\over 2g}\left\{
{1\over v} (\partial_t{\bf e})^2 - v (\partial_x{\bf e})^2 \right\}~~.
\end{equation}
The coupling $g$ and the velocity $v$ are related to the spin $s$, the
AF coupling $J$ and the lattice spacing: $v=2Jas$ and $g=2/s$.
To this action one must add a topological term $S_{top}$ for half-integer spin
(cf. Sec.~IV). An extension of this mapping to dimensions higher than one has
been obtained by many authors, with the difference that no topological term
arises, hence no distinction between half-integral and integral spin.

Dombre and Read\cite{Dombre89} have conducted a similar analysis for
the antiferromagnetic Heisenberg model on a triangular lattice. The
essential difference here is the presence of frustration, leading to a
classical ground state characterized by a $120^\circ$ order.
The local order must then be specified by a rotation matrix instead
of a unit vector and the order parameter is thus an element of
$SO(3)$. The long-wavelength action found in this case is the $SO(3)$
nonlinear sigma model, with Lagrangian density
\begin{equation}
\label{DR}
{\cal L} = {1\over 2g} \left\{ {1\over v}{\rm tr}(\partial_t R^{-1}
\partial_t R) -v{\rm tr}(P\nabla R^{-1}\cdot\nabla R) \right\}~~.
\end{equation}
Here $R({\bf x},t)$ is a position- and time-dependent rotation matrix and
$P$ is the constant diagonal matrix ${\rm diag}[1,1,0]$. Again, the constants
$g$ and $v$ depend on the spin $s$, the lattice spacing $a$ and the AF
couplings $J$. At zero temperature and in dimension two, this model has an
ordered phase for $g<g_c$ and a disordered phase otherwise. Since $g\sim a/s$,
the spin-$\frac12$ case is the closest to the disordered phase. However, it
is now widely believed\cite{noLRO} that the ground state of the spin-$\frac12$
antiferromagnet on a triangular lattice has long range order and
consequently the disordered phase of the $SO(3)$ model is not physically
realized to two-dimensional antiferromagnets, at least at half-filling (one
spin per site).

\begin{figure}[tpb]
\vglue 0.4cm\epsfxsize 8cm\centerline{\epsfbox{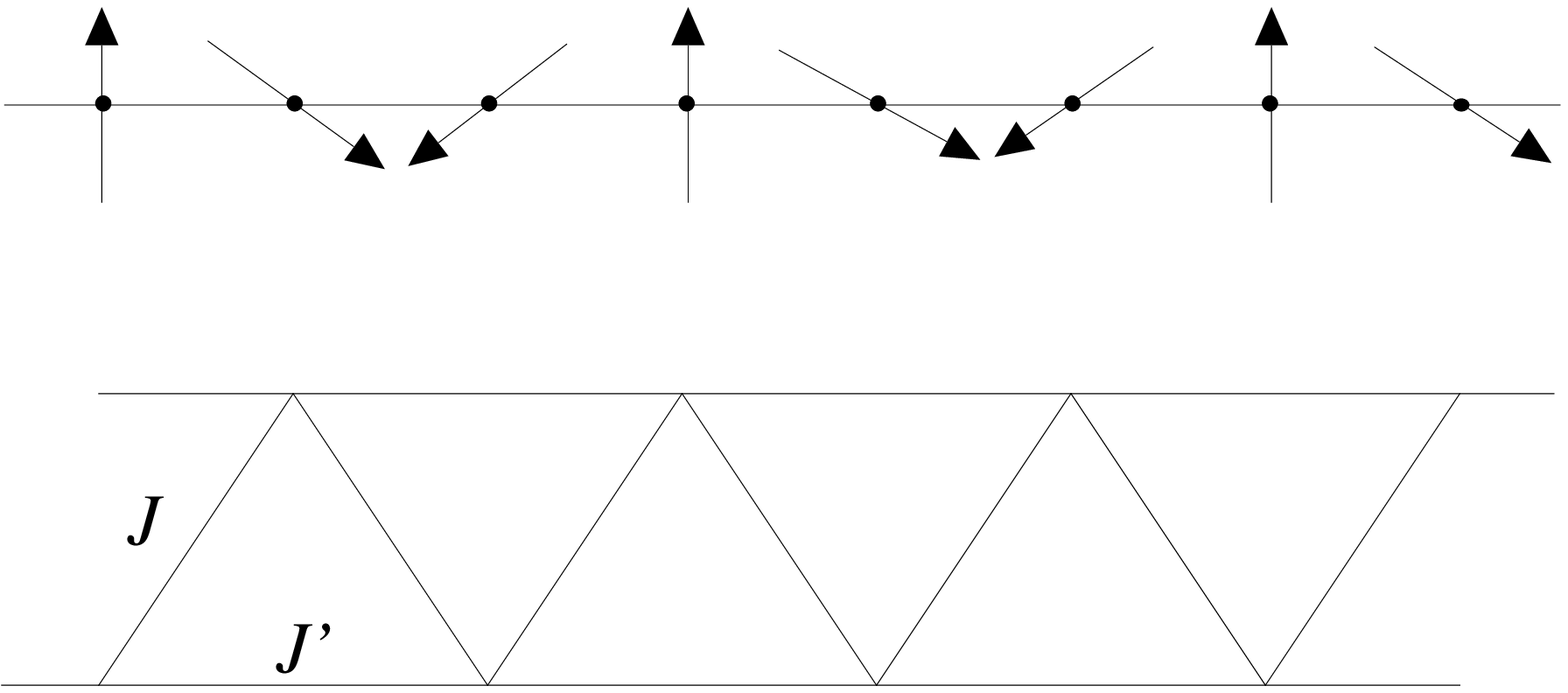}}\vglue 0.4cm
\caption{
Classical ground state of the frustrated AF chain with $J'/J=\frac12$.
Each spin makes a $120^\circ$ angle with its neighbors and the periodicity
is the smallest possible, i.e., three sites. Below is different look at the
same spin chain, using a `railroad trestle' geometry.
}
\label{groundFig}
\end{figure}

In this work we will argue that this disordered phase of the $SO(3)$
sigma model could be realized in a frustrated antiferromagnetic chain, with
Hamiltonian
\begin{equation}
\label{Hamil}
H = J\sum_n {\bf S}_n\cdot {\bf S}_{n+1} +
J'\sum_n {\bf S}_n\cdot {\bf S}_{n+2}~~.
\end{equation}
Introducing a next-nearest-neighbor (NNN) coupling $J'$ modifies the classical
ground state if $J'/J >\frac14$: the spins still lie on a common plane, but
instead of being collinear (as in the N\'eel state), they are arranged in a
canted configuration (see Fig.~\ref{groundFig}) in which each spin makes an
angle $\alpha$ with its predecessor, given by
\begin{equation}
\label{pitch}
\cos\alpha= -{J\over 4J'} ~~.
\end{equation}
In the special case $J'/J=\frac12$, the classical order is quite similar to
that of the triangular lattice, with its $120^\circ$ angle from site to site
and its periodicity of three sites. For spin-$\frac12$ this particular case
constitutes the Majumdar-Gosh model, whose ground state is exactly
known.\cite{Majumdar69} For a generic value of $J'/J$ the classical order is
incommensurable, with infinite periodicity. The spin-1 case has been studied
previously using various methods: exact diagonalizations\cite{Tonegawa92}, 
approximate mappings to field theories involving fermions\cite{Harada93} or
bosons\cite{Shimaoka93,Rao93}, among others.

This paper is organized as follows. In Sec.~II we explain how to obtain
the $SO(3)$ nonlinear sigma model as the continuum limit of the model
defined in Eq.~(\ref{Hamil}). In Sec.~III we argue that the disordered
phase of this model is characterized by a singlet ground state with a gap 
to {\it two} degenerate triplets of excited states. Hence this model cannot
adequately represent the half-integer NNN chain, but may be correct for
integer spin. In Sec.~IV we discuss our results with the help of exact
diagonalization results and discuss the apparent lack of distinction between
integer and half-integer spin in this model.

\section{Semi-classical mapping to the $SO(3)$ nonlinear sigma model}
%

In this section we argue that the long wavelength, continuum theory
describing the frustrated antiferromagnetic chain is specified by the
Lagrangian density of Eq.~(\ref{Lag}), with the notation explained
thereafter.

The Lagrangian description of spin dynamics, as used in path integrals,
requires the introduction of spin coherent states. The unfamiliar reader is
referred to Fradkin's text\cite{Fradkin91} or to Manousakis'
review\cite{Manousakis91}. Each quantum spin ${\bf S}_i$ is described by a
fluctuating unit vector ${\bf n}_i$ and the action associated to a spin
Hamiltonian $H$ is
\begin{equation}
\label{action1}
S = \int dt\left\{ s\sum_i ~A({\bf n}_i)\cdot\partial_t{\bf n}_i -H\right\}
\end{equation}
wherein ${\bf S}_i$ is replaced by $s{\bf n}_i$ in the Hamiltonian.
${\bf A}$ is the vector potential of a magnetic monopole of flux $4\pi$.
Thus, given a closed curve ${\bf n}(t)$ on the unit sphere, the 
integral $\int{\bf A}\cdot d{\bf n}$ equals the area of the sphere enclosed
by the curve, modulo $4\pi$.

We shall use a description of each spin ${\bf S}_i$ in terms of a slowly
varying $SO(3)$ order parameter $R(x,t)$ and a local magnetization vector
${\bf l}(x,t)$:
\begin{mathletters}
\label{rep}
\begin{eqnarray}
{\bf S}_k &=& s{R\left({\bf n}^c_k + a{\bf l}\right) 
\over |{\bf n}^c_k + a{\bf l}|} \\
{\bf n}_k^c &=& {\bf\hat\i}\cos(k\alpha) + {\bf\hat\j} \sin(k\alpha)
\end{eqnarray}
\end{mathletters}
Here ${\bf n}_k^c$ is the orientation of the spin ${\bf S}_k$ in a
classical ground state taken as reference, in which all spins lie on the
plane defined by two mutually orthogonal unit vectors ${\bf\hat\i}$ and 
${\bf\hat\j}$; $\alpha$ is the pitch of this ground state, as given in
Eq.~(\ref{pitch}); finally, $a$ is the lattice spacing: It is assumed that the
spin configurations that contribute significantly to the path integral are
locally close to the classical ground state, and the approximation that the
deviation $a{\bf l}$ is small will be controlled at the same time as the
continuum approximation.

Some comments are in order concerning this representation.
If $J'=J/2$ (or $\alpha=2\pi/3$), the periodicity is 3 and the relation
(\ref{rep}) may be considered as a {\it bona fide} change of variables
if we group the spins in sets of three and assume that the fields $R$
and ${\bf l}$ do not vary within such sets. We check that the number of
degrees of freedom match: $6$ per set of three spins. However, it is more
convenient to assume that $R$ and ${\bf l}$ vary slowly from site to site,
in which case the representation (\ref{rep}) may not be regarded as a
change of variables, but simply as long wavelength description of the
fluctuations around the classical ground state; then we need not restrict
ourselves to the case $J'=J/2$.

The next step is to substitute the representation (\ref{rep}) into the
action (\ref{action1}) and to Taylor expand when needed in order to get
a continuum action in terms of $R$ and ${\bf l}$ only. We will treat the
kinetic term first, and then the Hamiltonian. 
For the sake of convenience, we will assume a periodicity of $N$ spins
in the classical ground state. The case $N=3$ (or $J'=J/2$) is the simplest,
leading to a calculation almost identical to that of
Ref.~\onlinecite{Dombre89}. The next simplest is $N=5$, with 
$J'\approx 0.809J$, and so on. 


An explicit calculation of the kinetic term is not needed here, since
it would be almost identical to that of Ref.~\onlinecite{Dombre89}, to
which the reader is referred. The difference here lies in the number of
reference vectors ${\bf n}^c_i$, which is $N$ instead of 3. The kinetic part
of the Lagrangian density is then
\begin{equation}
\label{kinetic}
{\cal L}_K = as(T{\bf l})\cdot {\bf V}
\end{equation}
with the definitions
\begin{mathletters}
\begin{eqnarray}
T_{ab} &=& \delta_{ab} - \langle n^c_{i,a}n^c_{i,b}\rangle
 = \pmatrix{\frac12&0&0\cr 0&\frac12&0\cr0&0&0\cr}\\
V_a &=& -\frac12\epsilon_{abc}(R^{-1}\partial_t R)_{bc}
\end{eqnarray}
\end{mathletters}
wherein the indices in spin space correspond to the axes defined by 
${\bf\hat\i}$, ${\bf\hat\j}$ and ${\bf\hat\i}\times {\bf\hat\j}$, 
and $\langle\cdots\rangle$ means an average over $N$ contiguous sites.

In order to write down the continuum limit of the Hamiltonian $H$, we need
to calculate the interaction $H^{(k)}$ of a spin ${\bf S}_k$ with its four
neighbors and then average the result over the $N$ spins ${\bf S}_k$ of the
period. Specifically, the Hamiltonian density is
\begin{equation}
{\cal H} = {1\over 2aN}\sum_k H^{(k)} = {1\over 2a}\langle H^{(k)}\rangle~~.
\end{equation}
The contribution $H^{(k)}$ may be written as follows:
\begin{eqnarray}
H^{(k)} &=& J\left\{ 
{\bf S}_k\cdot{\bf S}_{k+1} + {\bf S}_k\cdot{\bf S}_{k-1}
+ {J'\over J} {\bf S}_k\cdot{\bf S}_{k+2}
+ {J'\over J} {\bf S}_k\cdot{\bf S}_{k-2} \right\} \nonumber\\
&+&aJ\left\{ 
 {\bf S}_k\cdot\partial_x{\bf S}_{k+1}
-{\bf S}_k\cdot\partial_x{\bf S}_{k-1}
+{2J'\over J}{\bf S}_k\cdot\partial_x{\bf S}_{k+2}
-{2J'\over J}{\bf S}_k\cdot\partial_x{\bf S}_{k-2} \right\} \nonumber\\
&+&{a^2J\over2}\left\{ 
 {\bf S}_k\cdot\partial^2_x{\bf S}_{k+1}
+{\bf S}_k\cdot\partial^2_x{\bf S}_{k-1}
+{4J'\over J}{\bf S}_k\cdot\partial^2_x{\bf S}_{k+2}
+{4J'\over J}{\bf S}_k\cdot\partial^2_x{\bf S}_{k-2} \right\} + \cdots
\end{eqnarray}
where the spin ${\bf S}_n$ stands in fact for its expression in terms of
$R$ and ${\bf l}$ (Eq.~(\ref{rep})) and where the derivatives act on these
same fields. We will carry the expansion to second order only, enough to
yield a non trivial Lagrangian. We shall also use a Taylor expanded version
of the representation (\ref{rep}):
\begin{equation}
{\bf S}_k = sR({\bf n}^c_k) + saR({\bf l}-[{\bf l}
\cdot {\bf n}^c_k]{\bf n}^c_k)   \nonumber\\
+ sa^2 R({\bf n}^c_k)\left[{\textstyle\frac32}({\bf n}^c_k\cdot{\bf l})^2-
{\textstyle\frac12}{\bf l}^2\right] -sa^2R({\bf l})[{\bf n}^c_k\cdot{\bf l}]
+{\cal O}(a^3)~~.
\end{equation}

We may then write the Hamiltonian density as an expansion in powers of $a$:
\begin{equation}
{\cal H}={\cal H}^{(0)}+a{\cal H}^{(1)}+a^2{\cal H}^{(2)}+\cdots
\end{equation}
${\cal H}^{(0)}$ is a constant, independent of $R$
and ${\bf l}$. ${\cal H}^{(1)}$ vanishes for the
following reasons: Part of it depends on ${\bf l}$, but it is proportional
to $\langle {\bf n}^c_k\rangle$, which is zero. Another part depends
on $R$, and is given by
\begin{equation}
\label{H1a}
{\cal H}^{(1)} = {s^2\over2a} (R^{-1}\partial_xR)_{ab}\Big\{
J(\langle n^c_{k,a}n^c_{k+1,b}\rangle 
- \langle n^c_{k,a}n^c_{k-1,b}\rangle)
+2J'(\langle n^c_{k,a}n^c_{k+2,b}\rangle 
- \langle n^c_{k,a}n^c_{k-2,b}\rangle) \Big\}~~.
\end{equation}
But one easily calculates that
\begin{equation}
\langle n^c_{k,a}n^c_{k+n,b}\rangle = 
\frac12\pmatrix{\phantom{-}\cos n\alpha & \sin n\alpha &0\cr 
-\sin n\alpha & \cos n\alpha &0 \cr 0&0&0\cr}~~.
\end{equation}
After substituting and using the relation (\ref{pitch}), we find out that the
expression (\ref{H1a}) vanishes.

Finally, we are left with the second order part ${\cal H}^{(2)}$. It is a
straightforward exercise to show that
\begin{equation}
{\cal H} = {Jas^2\over 2} l_a M_{ab} l_b - {Jas^2\over 4}\left(
\xi-{1\over\xi}\right) {\rm tr}[P\partial_xR^{-1}\partial_xR]
\end{equation}
where $\xi\equiv-\cos\alpha=J/4J'$ and where the diagonal matrix $M$ has
elements
\begin{equation}
M_{33} =  {(1+\xi)^2\over\xi}\qquad\qquad
M_{11}=M_{22} =  M_{33}(\xi^2-\xi+{\textstyle\frac12})~~.
\end{equation}
Again, the matrix $P$ is the projector onto the plane defined by 
${\bf\hat\i}$ and ${\bf\hat\j}$, as in Eq.~(\ref{DR}).
The full Lagrangian density is then ${\cal L}={\cal L}_K-{\cal H}$. 

The last step consists in integrating out the field ${\bf l}$.
Since the latter appears quadratically in ${\cal L}$, this operation amounts
to substituting into ${\cal L}$ the solution of the classical equations of
motion for ${\bf l}$. The final result for the Lagrangian density is
\begin{equation}
\label{Lag}
{\cal L} = {1\over 2\tilde g}\left\{ 
{1\over\tilde c}\,{\rm tr}[Q\partial_t R^{-1}\partial_t R]
-\tilde c\,{\rm tr}[P\partial_x R^{-1}\partial_x R]\right\}
\end{equation}
where the matrix $Q$ is diagonal, with elements
\begin{equation}
Q_{11}=Q_{22}=1 \qquad\qquad Q_{33}= {2\xi(1-\xi)\over \xi^2+(\xi-1)^2}
\end{equation}
and where the constants $\tilde g$ and $\tilde c$ are defined as
\begin{mathletters}
\begin{eqnarray}
\tilde g &=& {2\over s}\sqrt{1+\xi\over1-\xi}\\
\tilde c &=& Jas{1+\xi\over\xi}\sqrt{1-\xi^2}~~.
\end{eqnarray}
\end{mathletters}
In the special case $\xi=\frac12$, which corresponds to
$J'=J/2$ and $\alpha=2\pi/3$, one recovers precisely the form
of the Lagrangian (\ref{DR}) since $Q_{33}=1$. The parameter $\xi$ ranges
{}from 0 ($J'\to\infty$) to 1 ($J'=J/4$). The characteristic speed $\tilde c$
diverges as $\xi\to0$, whereas the coupling constant $\tilde g$ diverges as
$\xi\to1$, the collinear phase boundary. In all cases, the Lagrangian
(\ref{Lag}) does not possess Lorentz invariance, which makes it a
qualitatively distinct theory from the versions of the $SO(3)$ nonlinear
sigma model studied in the context of classical critical 
phenomena\cite{Azaria93}. Its symmetries comprise global left rotations
$R\to U_LR$ with $U_L\in SO(3)$ and global right rotations $R\to RU_R$ with
$U_R\in SO(2)$ (i.e. $U_R$ commuting with $P$ and $Q$). The first of these
reflects invariance under rotations in spin space, whereas the other will
have consequences on the spectrum of the theory: a degeneracy of the
excitation branches.

It is important to stress here that in going from the discrete Heisenberg
Hamiltonian (\ref{Hamil}) to the continuum Lagrangian (\ref{Lag}) we have
assumed that smooth configurations dominate the path integral.
There seems to be no distinction between integer and half-integer spins within
this mapping. In particular, there is no topological term of the type arising
in the collinear antiferromagnetic chain ($J'<J/4$). In fact, none could
exist, since the relevant homotopy group is trivial: $\pi_2(SO(3))=0$.
However, there are qualitative differences between integer and half-integer
spin in this system. The question as to why they do not appear explicitly in
the field theory will be discussed in Sec.~IV.

\section{Spectrum of the nonlinear sigma model}
In this section we discuss the spectrum of the model defined in
Eq.~(\ref{Lag}), in particular regarding the gap to excited states and
the degeneracy of the excitation branches. To this end it is preferable to
rewrite the Lagrangian~(\ref{Lag}) in terms of two orthonormal vectors ${\bf
e}_1$ and ${\bf e}_2$ specifying the rotation matrix $R$. Defining 
${\bf e}_3={\bf e}_1\times{\bf e}_2$, the elements of the matrix $R$ may be
expressed as $R_{ab}=({\bf e}_b)_a$ and ${\rm tr}[Q\partial_t
R^{-1}\partial_tR] =\sum_a Q_{aa}(\partial_t{\bf e}_a)^2$. 
Using the fact that
\begin{equation}
(\partial_t{\bf e}_3)^2 = (\partial_t{\bf e}_1)^2 + (\partial_t{\bf e}_2)^2
-2({\bf e}_1\cdot\partial_t{\bf e}_2)^2~~,
\end{equation}
we may finally express the Lagrangian in the following form:
\begin{equation}
\label{LagVec}
{\cal L} = {1\over2g}\left\{ 
{1\over c}\left[(\partial_t{\bf e}_1)^2 + (\partial_t{\bf e}_2)^2\right]
-{\gamma\over c}({\bf e}_1\cdot\partial_t{\bf e}_2)^2
-c\left[(\partial_x{\bf e}_1)^2 + (\partial_x{\bf e}_2)^2\right]\right\}
\end{equation}
wherein the coupling constants $g$, $\gamma$ and the velocity $c$ are now
\begin{mathletters}
\label{params}
\begin{eqnarray}
g &=& {2\over s}\sqrt{(1+\xi)(2\xi^2-2\xi+1)\over1-\xi} \\ 
c &=& Jas{1+\xi\over\xi}\sqrt{(1-\xi^2)(2\xi^2-2\xi+1)} \\
\gamma &=& 4\xi(1-\xi)
\end{eqnarray}
\end{mathletters}
The functional integration measure associated with the fields ${\bf e}_{1,2}$
must incorporate the constraints ${\bf e}_1^2={\bf e}_2^2=1$,
${\bf e}_1\cdot{\bf e}_2=0$. This may be done by introducing three Lagrange
parameters $\sigma_{11}$, $\sigma_{22}$ and $\sigma_{12}=\sigma_{21}$, and by
adding the following constraint term to the Lagrangian:
\begin{equation}
{\cal L}_{constr} = 
\textstyle\frac12 \sigma_{ab}({\bf e}_a\cdot{\bf e}_b-\delta_{ab})~~.
\end{equation}

Since this model is defined in one spatial dimension, the
Mermin-Wagner-Coleman theorem applies and the global $SO(3)$
symmetry is not broken. If we extended the model to dimensions greater than
one, a broken symmetry phase could be realized, for $g$ below some critical
value. In such a phase, three Goldstone modes would appear, corresponding to
the three parameters of the broken $SO(3)$ symmetry. These three Goldstone
modes would have a linear dispersion relation, two of them with speed $c$
and a third with speed $c/\sqrt{1-\gamma/2}$. The most elegant way to see
this is to go back to the form (\ref{Lag}) of the Lagrangian and to 
substitute, in the small oscillation approximation,
$R_{ab}\approx\delta_{ab}+\varepsilon_{abc}\Omega_c$, keeping the terms
of order $\Omega_c^2$ only. Since the field $\Omega_c$ is unconstrained,
the dispersion relations are easily read off. In the case of an
antiferromagnet on a triangular or hexagonal lattice ($\xi=\frac12$), the
two speeds are $c$ and $c\sqrt2$. This may also be observed within 
spin-wave theory, since it is a feature of the ordered phase.

However, we are interested here in the disordered phase of the model,
which we shall study within the large-$N$ approach (here $N$ is the
number of components of the vectors ${\bf e}_a$, normally 3).
In the imaginary-time formalism, the partition function of the model is
\begin{equation}
Z = \int[d{\bf e}_1][d{\bf e}_2][d\sigma_{ab}] \exp -\int dxd\tau~{\cal L}_E
\end{equation}
with the Euclidian Lagrangian density
\begin{equation}
{\cal L}_E = \frac12\left\{ 
\left[(\partial_\tau{\bf e}_1)^2 + (\partial_\tau{\bf e}_2)^2\right]
-g\gamma({\bf e}_1\cdot\partial_\tau{\bf e}_2)^2
+\left[(\partial_x{\bf e}_1)^2 + (\partial_x{\bf e}_2)^2\right]\right\}
+\sigma_{ab}({\bf e}_a\cdot{\bf e}_b-{1\over g}\delta_{ab})
\end{equation}
where we have rescaled ${\bf e}_a$ by a factor $\sqrt g$ in order to
recover the standard normalization for the kinetic term. The characteristic
speed $c$ has been set to unity in order to lighten the notation; it may
be restored by dimensional analysis. We will also
use a Hubbard-Stratonovich decomposition of the quartic term:
\begin{equation}
\exp-\int dxd\tau~{g\gamma\over2}({\bf e}_1\cdot\partial_\tau{\bf e}_2)^2
=\int[d\phi]\exp-\int dxd\tau~\left\{
\textstyle\frac12\phi^2 
-\sqrt{g\gamma}\phi({\bf e}_1\cdot\partial_\tau{\bf e}_2\right\})
\end{equation}

We then proceed to find the large-$N$ saddle point. In other words, we
assume that the auxiliary fields $\phi$ and $\sigma_{ab}$ take a constant
value, which is determined by extremizing the effective potential obtained
by integrating the fields ${\bf e}_a$. This exercise is better done in
Fourier space, in which the Euclidian action with constant auxiliary fields
may be written as
\begin{mathletters}
\label{Lag2}
\begin{equation}
S_E = \frac12\int{d\omega\over2\pi}{dk\over2\pi}~
{\bf e}^*_a(\omega,k)K_{ab}{\bf e}_b(\omega,k) 
+{\textstyle\frac12} L^2\left[\phi^2 
- {1\over g}\sigma_{ab}\delta_{ab}\right]
\end{equation}
\begin{equation}
K(\omega,k) = \pmatrix{ 
\omega^2+k^2+\sigma_{11} & \sigma_{12} -i\omega\phi\sqrt{g\gamma} \cr
\sigma_{12} -i\omega\phi\sqrt{g\gamma} & \omega^2+k^2+\sigma_{22} \cr}
\end{equation}
\end{mathletters}
The effective potential is then
\begin{equation}
V_{eff} = {\textstyle\frac12}\phi^2 -{1\over2g}(\sigma_{11}+\sigma_{22})
+{N\over2}\int{d\omega\over2\pi}{dk\over2\pi}~\log\det K(\omega,k)
\end{equation}
where we assume the system to be limited by a box of side $L$.

In terms of the variables 
\begin{equation}
\sigma\equiv{\textstyle\frac12}(\sigma_{11}+\sigma_{22})
\qquad{\rm and}\qquad
\eta^2\equiv {\textstyle\frac14}(\sigma_{11}-\sigma_{22})^2 + \sigma_{12}^2
\end{equation}
the saddle-point equations are the following:
\begin{mathletters}
\begin{eqnarray}
{\partial V_{eff}\over\partial\sigma} &=& 0 =
-{1\over g} + {N\over2}\int{d\omega\over2\pi}{dk\over2\pi}~
{2(\omega^2+k^2+\sigma)\over(\omega^2+k^2+\sigma)^2-\eta^2-
g\gamma\phi^2\omega^2} \\
{\partial V_{eff}\over\partial\eta} &=& 0 =
N\eta\int{d\omega\over2\pi}{dk\over2\pi}~
{1\over(\omega^2+k^2+\sigma)^2-\eta^2- g\gamma\phi^2\omega^2} \\
{\partial V_{eff}\over\partial\phi} &=& 0 =
\phi - N\phi g\gamma\int{d\omega\over2\pi}{dk\over2\pi}~
{\omega^2\over(\omega^2+k^2+\sigma)^2-\eta^2-g\gamma\phi^2\omega^2}
\end{eqnarray}
\end{mathletters}
The immediate solution to these equations is $\phi=\eta=0$ and $\sigma\ne0$,
determined by the simpler equation
\begin{equation}
\label{gapEq}
{1\over g} = {N\over2}\int{d\omega\over2\pi}{dk\over2\pi}~
{1\over \omega^2+k^2+\sigma}
\end{equation}
It can be shown explicitly\cite{Allen94} that the solution $\phi=\eta=0$ is
the only acceptable one in this case.
Eq.~(\ref{gapEq}) coincides with the saddle-point equation for the
$O(3)/O(2)$ nonlinear sigma model, whose solution is
\begin{equation}
\sigma = \sigma_0 \equiv \Lambda^2\exp-{8\pi\over3g} 
\end{equation}
wherein $\Lambda$ is a momentum cutoff. The fields $\sigma_{11}$ and
$\sigma_{22}$ must therefore be shifted by the constant $\sigma_0$ in order
to fluctuate about zero. Then $\sigma_0$ multiplies
${\textstyle\frac12}{\bf e}_1^2$ and ${\textstyle\frac12}{\bf e}_2^2$
in the large-$N$ effective action; thus, it has the interpretation of a
quantum fluctuation-induced mass squared for the fields ${\bf e}_a$.
The excitations of these fields are then triplets with energy gap
$\Delta=\Lambda\exp-(4\pi/3g)$. After restoring the characteristic
speed $c$, and the dependence on $\xi$, this becomes
\begin{equation}
\Delta = (\Lambda a) Js{1+\xi\over\xi}\sqrt{(1-\xi^2)(2\xi^2-2\xi+1)}~
\exp-\left\{{2\pi s\over3}\sqrt{1-\xi\over(1+\xi)(2\xi^2-2\xi+1)}\right\}
\end{equation}
Of course, this result neglects $1/N$ corrections. As $\xi\to1$ (the
collinear point) the gap goes to zero. This signals a continuous
phase transition, beyond which a different description of the system is
needed, for instance in terms of the $O(3)/O(2)$ sigma model.
{}From $\xi=1$ up to $\xi=0$, which corresponds to $J'\to\infty$, the
gap increases and becomes eventually proportional to $J'$.

The main feature of the excitations of this model is the degeneracy
of the triplets. This ultimately comes from the $SO(2)$ symmetry
existing between the vectors ${\bf e}_1$ and ${\bf e}_2$, which may
be mixed with each other without affecting the action. It is then not
surprising if this degeneracy survives the large-$N$ approximation.

\begin{figure}[tpb]
\vglue 0.4cm\epsfxsize 6cm\centerline{\epsfbox{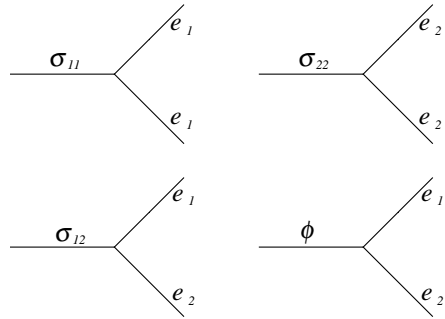}}\vglue 0.4cm
\caption{
Vertices used in the $1/N$ expansion of the model (27).
}
\label{vertices}
\end{figure}

\begin{figure}[tpb]
\vglue 0.4cm\epsfxsize 8cm\centerline{\epsfbox{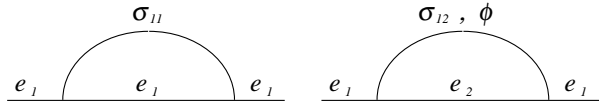}}\vglue 0.4cm
\caption{
$1/N$ contributions to the self-energy $\Sigma_{11}$ of the 
field ${\bf e}_1$.
}
\label{self}
\end{figure}

Indeed, $1/N$ corrections to the dispersion of the excitations can in
principle be calculated, like they are in the $O(3)/O(2)$ sigma model
(see, for instance, Polyakov's text\cite{Polyakov87}). The quantity of
interest is then the self-energy of the fields ${\bf e}_a$. In general,
this is a matrix $\Sigma_{ab}(\omega,k)$, which may be calculated
diagrammatically with the help of the vertices shown in Fig.~\ref{vertices}
and of the propagators for the auxiliary fields $\sigma_{ab}$ and $\phi$. The
latter scale as $1/N$ since they are inverse polarization operators, and the
order in $1/N$ is simply determined by the number of auxiliary field internal
lines. However, the propagators for $\sigma_{11}$ and $\sigma_{22}$ are
identical, since the large-$N$ propagators of ${\bf e}_1$ and ${\bf e}_2$ 
are the same. Moreover, the structure of the vertices of Fig.~\ref{vertices}
makes it impossible for a non-diagonal self-energy $\Sigma_{12}$ to arise:
even though the auxiliary fields $\sigma_{12}$ and $\phi$ may change a ${\bf
e}_1$ quanta into a ${\bf e}_2$ quanta, the number of such vertices must be
even, without tadpoles. The net result is that no self-energy diagram exists
in which an entering ${\bf e}_1$ line is turned into an exiting ${\bf e}_2$
line, and consequently $\Sigma_{12}=0$. The $1/N$ contributions to the
$\Sigma_{11}$ are show diagrammatically on Fig.~\ref{self}. At all orders in
$1/N$ the two self-energies $\Sigma_{11}$ and $\Sigma_{22}$ are identical and
therefore the degeneracy of the excitations branches seems a exact feature of
the model.

\section{Discussion}

In Sec~II we argued that the frustrated AF spin chain may be described,
in the continuum limit, by the $SO(3)$ nonlinear sigma model defined
by the Lagrangian (\ref{Lag}). In Sec~III, we argued that the main feature of
the spectrum of this continuum model is a doubly degenerate triplet of spin-1
excitations, separated from the singlet ground-state by a gap. However, this
conclusion is incompatible with what is known of the spectrum of the
frustrated spin-$\frac12$ chain. The Lieb-Schultz-Mattis (LSM)
theorem\cite{Lieb61} states that a spin-$\frac12$ chain has either no gap
or degenerate ground states (corresponding to spontaneously broken parity).
More recently, it has been shown\cite{Affleck86} that this theorem applies to
half-integer spin chains in general. In the special case $J'=J/2$ the exact
ground state of the spin-$\frac12$ chain obtained by Majumdar and
Gosh\cite{Majumdar69} is two-fold degenerate. This is compatible with the LSM
theorem, but in contradiction with the field theory. Thus, the latter
does not correctly describe frustrated half-integer spin chains.

\begin{figure}[tpb]
\vglue 0.4cm\epsfxsize 8cm\centerline{\epsfbox{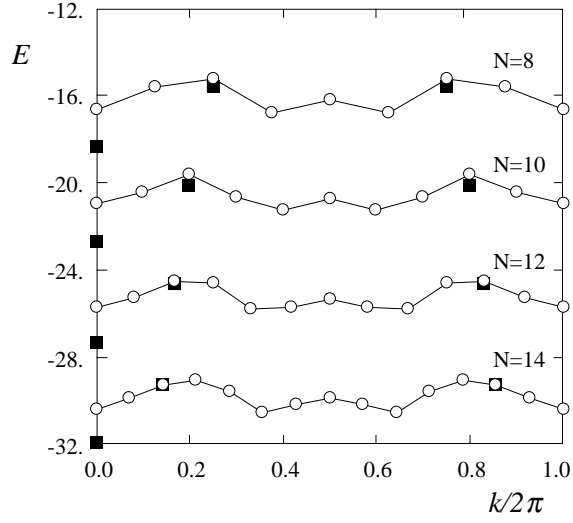}}\vglue 0.4cm
\caption{
Excitation spectrum of the frustrated spin-1 chain with $J'=J/2$ for chains
lengths $L=8$, 10, 12 and 14. The white circles represent spin-triplet states,
while the black squares represent spin-singlets, when lower in energy.
}
\label{spectre1}
\end{figure}

\begin{figure}[tpb]
\vglue 0.4cm\epsfxsize 8cm\centerline{\epsfbox{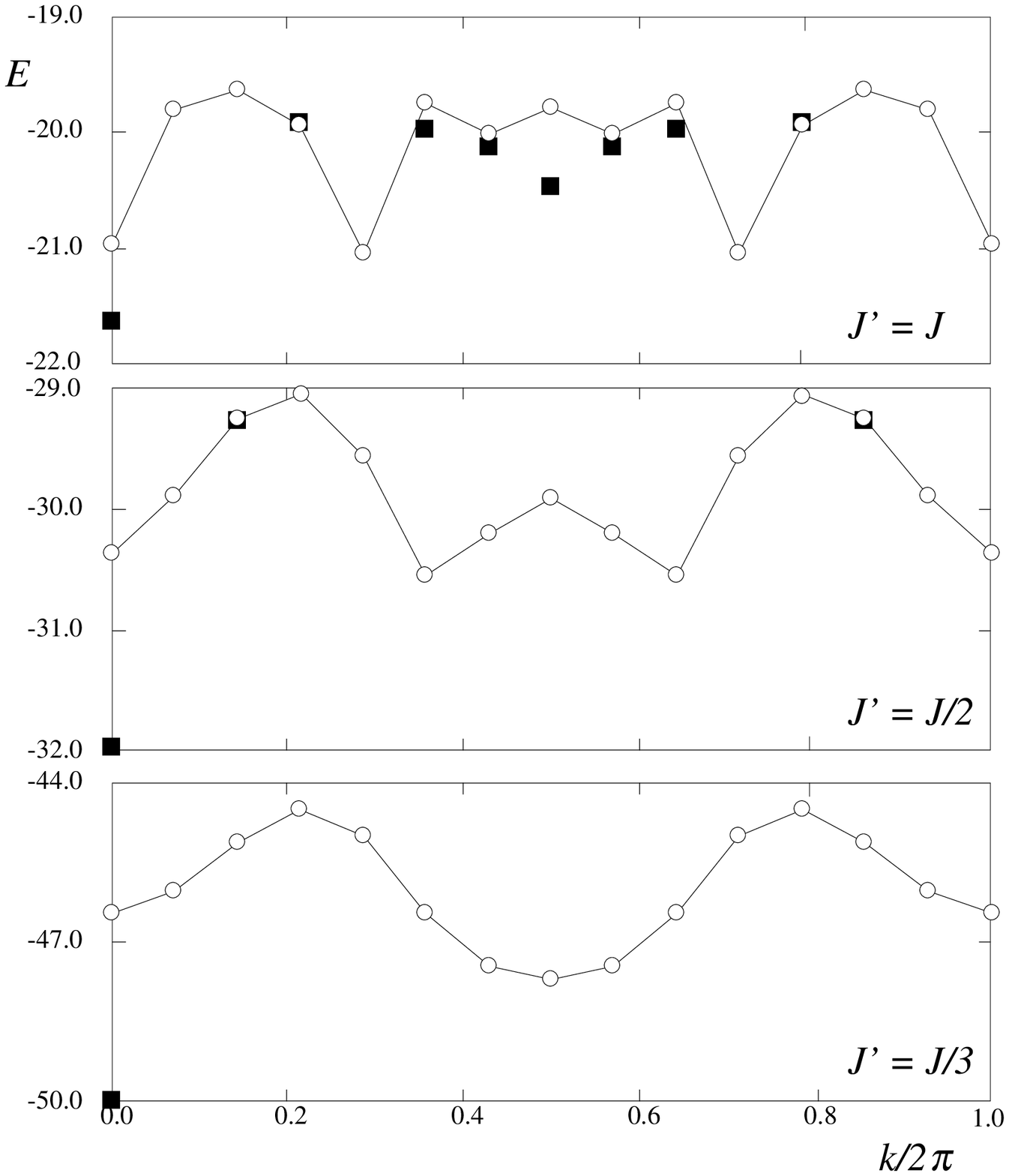}}\vglue 0.4cm
\caption{
Excitation spectrum of the 14-site frustrated spin-1 chain with $J'/J=1$,
$\frac12$ and $\frac13$. Again, the white circles represent spin-triplet
states and the black squares spin-singlets, when lower in energy.
}
\label{spectre2}
\end{figure}

Before throwing the $SO(3)$ model away, one should see if it describes
the long wavelength behavior of integer spin chains. Exact results for the
spin-1 chain are lacking and thus we are reduced to numerical study of finite
systems. Quantum Monte-Carlo studies of frustrated spin systems are quite
difficult because of a sign problem, therefore we limited ourselves to
numerical diagonalizations of small chains (up to 14 sites). We used various
methods, of which the most practical turned out to be a variant of the
Lanczos method\cite{Gagliano86} applied to subspaces of fixed momentum.
Diagonalizations of spin-$\frac12$ chains with up to 22 sites have also been
performed, for the sake of comparison. It turns out that the spectra of
spin-1 and spin-$\frac12$ chains of small lengths are qualitatively different.
The ground state of the spin-1 chains is non-degenerate, with a gap to a
doubly degenerate triplet of spin-1 states, exactly as in the field theory.
Excitation spectra for $J'=J/2$ and various chain lengths are shown in
Fig.~\ref{spectre1}, whereas spectra for $14$ spins and various values
of $J'/J$ are shown in Fig.~\ref{spectre2}. The dispersion relation is
symmetric with respect to parity ($k\to-k$) and the minimum of the spin-1
branch occurs at a wavenumber $k_0$ smaller than $\pi$ (in units
of $1/a$). Consequently, this minimum also occurs at $-k_0$, which is
distinct from $k_0$ except at the ferromagnetic ($k=0$) and
antiferromagnetic ($k=\pi$) points. The existence of these minima away from
the parity-invariant points should not surprise us. If it were not for the
quantum fluctuation disordering the system, there would be massless Goldstone
branches around the ordering wave-vectors $k_0=\alpha$ and $-k_0$ ($\alpha$
is given by Eq.~(\ref{pitch})). The effect of quantum fluctuations is to
produce a singlet ground state and to raise the minima of the Goldstone
branches. We expect that the positions of these minima be approximately the
same as the classical ordering wave-vectors $k_0$ and $-k_0$, although we
have no way to prove that they coincide exactly with these values.
The $SO(3)$ field theory effectively describes low-energy excitations about
the minima of these excitation branches, and the degeneracy is essentially
linked to the fact that the classical ordering wavevectors $k_0$ and $-k_0$
are inequivalent. This feature of the excitation spectrum supports the view
that the $SO(3)$ field theory describes the long wavelength behavior of the
frustrated spin-1 chain.

\begin{figure}[tpb]
\vglue 0.4cm\epsfxsize 8cm\centerline{\epsfbox{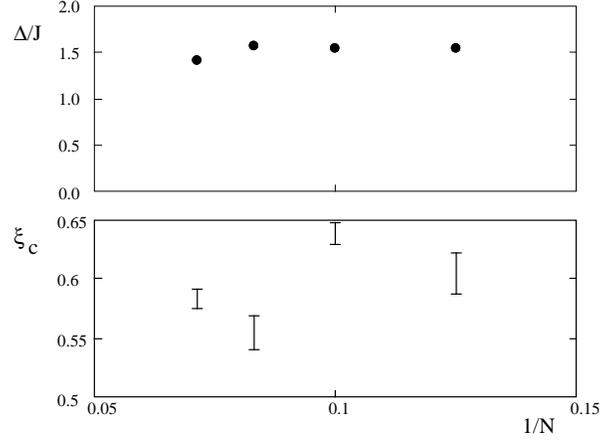}}\vglue 0.4cm
\caption{
Dependence on chain length $N$ of the critical coupling ratio
$\xi_c = (J/4J')_c$ at which the triplet degeneracy appears (error bars).
Also illustrated is the energy gap $\Delta/J$ for $\xi=\frac12$ (circles).
}
\label{spectre3}
\end{figure}

It should be pointed out that, in a chain of length $N$, there is critical
value $\xi_c^{(N)}$ of the ratio $J/4J'$, above which the minimum in the
triplet dispersion relation occurs at $k=\pi$. This critical value depends on
the length of the chain. If one trusts the heuristic mapping of section II,
it should be equal to $\xi_c^{(\infty)}=1$ when $N\to\infty$. For finite
chains it is closer to $\frac58$, but it is conceivable that it reaches
the predicted value as $N$ grows. Figure~\ref{spectre3} shows the evolution
of $\xi_c$ for small sizes, as well as the value of the mass gap $\Delta$.

The question remains as to why this semi-classical treatment fails in the
half-integer spin case. In other words, where in the argument of Sec.~II
should the distinction between integer and half-integer spin come into play? 
The answer may lie in our neglect of configurations
with discontinuities. This has already been pointed out in
Ref.~\onlinecite{Dombre89} in relation with the kinetic term. An easier
way to see this is to consider the limit $J'>J$. In that limit
one may reconsider the problem and view the system as a pair of interwoven,
antiferromagnetic chains with a small interchain interaction $J$. Each
chain could be semi-classically described by the usual $O(3)/O(2)$ nonlinear
sigma model, with the Lagrangian density (\ref{NLs}), plus a topological term
${\cal L}_{top}$ expressed as
\begin{equation}
{\cal L}_{top} = -{s\over2}\;{\bf e}\cdot
(\partial_t{\bf e}\times\partial_x{\bf e})~~.
\end{equation}
Since the effective lattice spacing for each chain is now $\tilde a=2a$, the
characteristic velocity would be $c=4J'as$, while the coupling constant is
still $g=2/s$. One checks that this agrees perfectly with the
$\xi\to0$ limit of Eq.~(\ref{params}). Thus, as $\xi=J/4J'$ decreases, the
two unit vectors ${\bf e}_1$ and ${\bf e}_2$ of the $SO(3)$ model become the
fields of two weakly coupled $O(3)/O(2)$ sigma models, with the correct
values of $c$ and $g$.

The interaction ${\cal L}_{int}$ between the two chains may be easily
expressed in terms of ${\bf e}_1$ and ${\bf e}_2$. Indeed, recall that
in taking the continuum limit of a single AF chain, one uses the
decomposition ${\bf S}_i=s[(-1)^i{\bf e}+\tilde a{\bf l}]$, where ${\bf l}$ is
the local magnetization and $\tilde a=2a$ is the effective lattice spacing for
each chain. Neglecting derivatives, the interaction Lagrangian may be written
as
\begin{equation}
{\cal L}_{int} ~=~ -2Js^2\tilde a\; {\bf l}_1\cdot{\bf l}_2~~.
\end{equation}
The elimination of ${\bf l}_1$ and ${\bf l}_2$ is done by substituting the
equation of motion\cite{Fradkin91}
\begin{equation}
{\bf l}_a = -{1\over 4J's\tilde a}({\bf e}_a\times\partial_t{\bf e}_a)~~.
\end{equation}
The interaction Lagrangian then becomes
\begin{eqnarray}
{\cal L}_{int} &=& -{1\over 16aJ'}{J\over J'}
({\bf e}_1\times\partial_t{\bf e}_1)\cdot
({\bf e}_2\times\partial_t{\bf e}_2) \\
 &=& -{1\over aJ}\xi^2\left\{
{\bf e}_1\cdot{\bf e}_2\; \partial_t{\bf e}_1\cdot\partial_t{\bf e}_2 -
({\bf e}_1\cdot\partial_t{\bf e}_2)
({\bf e}_2\cdot\partial_t{\bf e}_1)\right\}~~.
\end{eqnarray}
If we assume that the two vectors ${\bf e}_1$ and ${\bf e}_2$ are identically
orthogonal, as the classical ground state suggests if $\xi$ is small, then
the above interaction exactly agrees with Eq.~(\ref{LagVec}) and the
$\xi\to0$ limit of $\gamma$ as given in Eq.~(\ref{params}). Moreover, the
topological terms $S_{top}[{\bf e}_1]$ and $S_{top}[{\bf e}_2]$ cancel each
other if ${\bf e}_1$ is identically orthogonal to ${\bf e}_2$. This is
particularly easy to understand if the configuration ${\bf e}_1$ belongs to
the homotopy class of the identity, since the orthogonality
constraint then effectively makes ${\bf e}_2$ a mapping from the sphere $S_2$
to the circle $S_1$ and $\pi_2(S_1)=0$. In general, the two topological
terms cancel simply because of the fact that two orthogonal unit vectors
specify a rotation and $\pi_2(SO(3))=0$. Thus, if we assume that
${\bf e}_1\perp{\bf e}_2$ in the small $\xi$ limit, the Lagrangian
(\ref{LagVec}) is recovered with exactly the same parameters, since the
topological terms are absent.

However, the orthogonality ${\bf e}_1\perp{\bf e}_2$ is not strict, but
only favored energetically. A local deviation from this orthogonality gives
back their full importance to the topological terms and causes a distinction
between integer and half-integer spins. Such a distinction does not occur in
the $SO(3)$ formulation of the problem since the orthogonality is 
then `built-in'. In that formulation, a local deviation from
orthogonality corresponds to a discontinuity in the order parameter, which
was not allowed from the start.

Finally, let us point out that the above remarks concerning parity-breaking
ordering wavevectors in frustrated antiferromagnets and their consequences on
the excitation spectrum may apply to other systems, such as the
nearest-neighbor Heisenberg antiferromagnet on a triangular lattice. In this
case it is believed that the ground state has long-range order. However, it is
conceivable that the introduction of vacancies destroys this order without
affecting the applicability of the $SO(3)$ theory: the effect of the
vacancies would then be to increase the value of the effective coupling
constant $g$ beyond the critical value $g_c$. A priori, it is not clear if
the $SO(3)$ field theory describes the long wavelength behavior of
antiferromagnets on the Kagom\'e lattice or on the see-saw
chain\cite{Long91}. A semi-classical study of these systems may be
interesting.

\acknowledgements
The authors thank L. Chen, A. Chubukov, A.-M.~S. Tremblay and G. Zumbach for
stimulating discussions. Financial support from the Natural Sciences and
Engineering Research Council of Canada (NSERC) and le Fonds pour la Formation
de Chercheurs et l'Aide \`a la Recherche du Gouvernement du Qu\'ebec (FCAR) is
gratefully acknowledged.



\input epsf.tex
\end{document}